# Chaotic and non-chaotic mixed oscillations in a logistic systems with delay

Marek Berezowski[a, b,] Artur Grabski[b]

[a] Institute of Chemical Engineering and Physical Chemistry, Cracow University of Technology, 31-155 Krakow, ul. Warszawska 24, Poland

[b] Institute of Chemical Engineering, Polish Academy of Sciences, 44-100 Gliwice, ul. Baltycka 5, Poland

## Abstract

The paper deals with the theoretical analysis of a logistic system composed of at least two elements with distributed parameters. It has been shown that such a system may generate specific oscillations in spite of the fact that the solutions of the mathematical method are characterized by no dynamic bifurcations. It has also been shown that the time series of the state variables of such a system may behave in a semi-chaotic way. This means that they have then predictable and unpredictable fragments. The analysis has been illustrated by two examples, viz. of a simple logistic model and of a reactor with feedback.

## 1. Introduction

Based on the usual general definition it is customary to assume that chaos is uniquely unpredictable. The time series are totally irregular, the Poincaré sections yield sets of points scattered on a plane, and the Lyapunov exponents are positive. This definition, however, cannot be always confirmed. For instance, there exist systems which are able to generate signals, being chaotic and repeatable at the same time. We shall name them semi-chaotic signals. As it will be shown later, they are generated by devices capable of generating oscillations despite the lack of any bifurcation. These devices will be called flip-flop states.

## 2. Theoretical principles

The principle of generation of the flip-flop oscillations, mentioned above, is simple. Let us take into account a model built of two (or more) coupled elements with distributed parameters, i.e. elements introducing delay to a given system (Fig. 1). Let us assume that steady-state solutions of such a system are characterized by a multiplicity of states. The latter may be stable or unstable (Fig. 2). Let us further assume that none of these solutions generates autonomous oscillations, which means that it is not connected with any dynamic bifurcation. By introduction of $x$ and $y$ variables in the proximity of the states $L$ and $U$, the system will generate specific oscillations consisting in jumps of the values of these variables between individual steady-state solutions. This means that the variables $x_L$ and $y_L$ will periodically assume values corresponding to the lower steady state, whereas the variables $x_U$ and $y_U$ will periodically assume values corresponding to the upper steady state. This results from the fact that the variable $x$ forces the state of the variable $y$ and the latter variable $y$ forces the state of the variable $x$. The period of these oscillations is equal to $t_x+t_y$ (Fig. 3). Let us assume now that one of the steady states (e.g. the upper one) generates autonomous oscillations, resulting from a dynamic bifurcation [1]. This means that the variables $x_U$ and $y_U$ will also vary in an oscillatory manner, according to the mentioned bifurcation. In other words, the variables $x_U$ and $y_U$ will be modulated. Let us assume in turn that these autonomous oscillations are of chaotic type. The variables $x_U$ and $y_U$ will then also vary chaotically. In consequence we obtain semi-chaotic time series of the variables $x$ and $y$. The concrete solutions will be exemplified in further parts of this paper.

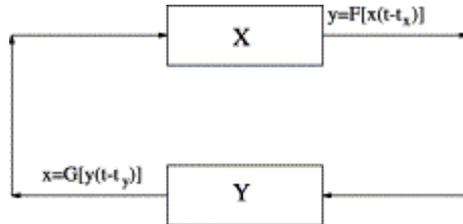

Fig. 1. Conceptual scheme of a flip-flop state.

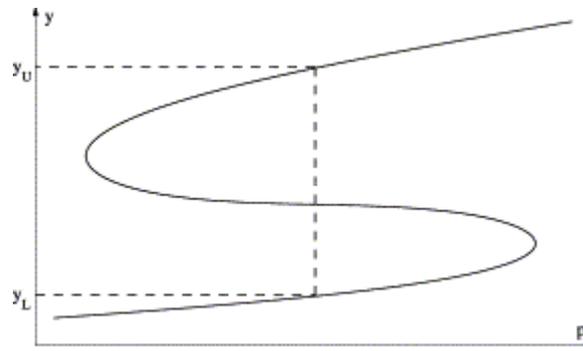

Fig. 2. Conceptual bifurcation diagram of a flip-flop state.

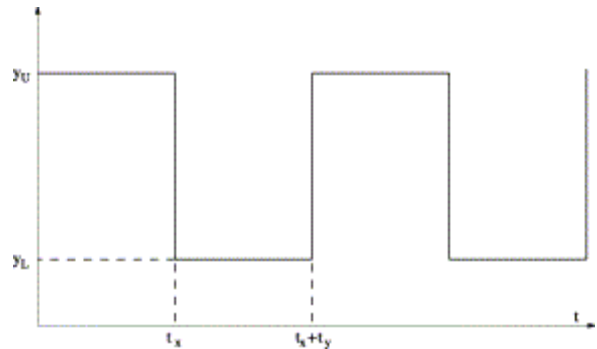

Fig. 3. Illustration of flip-flop oscillations.

## 3. Simple logistic model with delay

Let us take into consideration the logistic model [2] and [3]

$$y_{k+1} = ax_k(1 - x_k),$$ (1a)
$$x_{k+1} = y_k.$$ (1b)

Let us assume that the value $x_k$ is transmitted to $y_{k+1}$ with a certain arbitrary delay $t_x$ and that the value $y_k$ is transmitted to $x_{k+1}$ with a certain arbitrary delay $t_y$ (Fig. 1). Hence, such a model may be represented as [3]

$$y(t) = ax(t - t_x)\{1 - x(t - t_x)\}, \tag{2a}$$
$$x(t) = y(t - t_y). \tag{2b}$$

The system (1a)–(2b) has two steady states, $x_{s1}=y_{s1}=0$ and $x_{s2}=y_{s2}=(a-1)/a$. On the other hand, for $1<a<4$, the state $x_{s2}=y_{s2}$ can generate autonomous oscillations, including chaos. Let us assume, for example, the value $a=2$. In this case the trajectories should tend only and solely to the value $x=y=0.5$. This would be true if not for the delay $t_y$. Namely, it is sufficient to initiate the calculations of the system (1a)–(2b) at the initial conditions $x=0$ and $0<y<1$ (or $0<x<1$ and $y=0$) to obtain flip-flop oscillations as in Fig. 4. Let us, in turn, assume the value $a=3.8$. In this case $x_{s2}=y_{s2}$ is characterized by chaotic oscillations. Starting with the system (1a)–(2b) from $x=1$ and $0<y<1$ (or $0<x<1$ and $y=0$) one obtains the periodical time series of lower values of the variables $x$ and $y$, and chaotic time series of upper values of this variables. Hence, in this semi-chaotic time series the bottom is predictable, whereas the top is not ( Fig. 5).

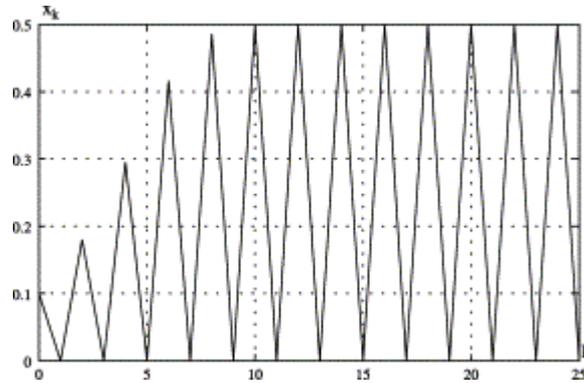

Fig. 4. Flip-flop oscillations of a simple logistic model ($a=2$).

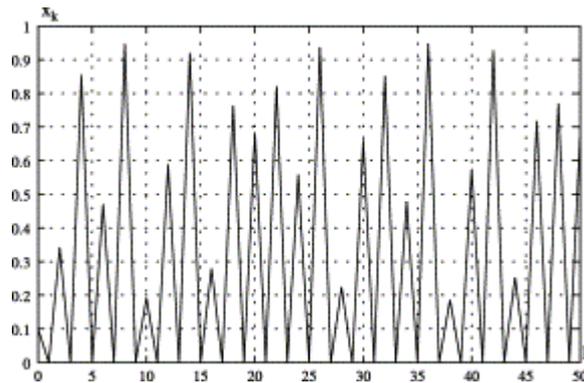

Fig. 5. Semi-chaotic flip-flop oscillations of a simple logistic model ($a=3.8$).

# 4. Model of a reactor with feedback

The model of a reactor fulfils the conditions of the logistic model described in <u>Section 1</u> (<u>Fig. 6</u>). The mathematical description of this model looks as follows.

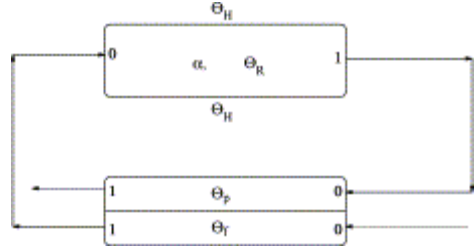

Fig. 6. Scheme of a reactor with feedback.

*Balance equations*:

$$\frac{\partial \alpha}{\partial \tau} + \frac{\partial \alpha}{\partial \xi_R} = \phi(\alpha, \Theta_R), \tag{3}$$

$$\frac{\partial \Theta_R}{\partial \tau} + \frac{\partial \Theta_R}{\partial \xi_R} = \phi(\alpha, \Theta_R) + \delta(\Theta_H - \Theta_R), \tag{4}$$

$$\sigma \frac{\partial \Theta_f}{\partial \tau} + \frac{\partial \Theta_f}{\partial \xi_w} = \delta_w(\Theta_p - \Theta_f), \tag{5}$$

$$\sigma \frac{\partial \Theta_p}{\partial \tau} + \frac{\partial \Theta_p}{\partial \xi_w} = -\delta_w(\Theta_p - \Theta_f). \tag{6}$$

*Boundary conditions*:

$$\alpha(0) = 0; \qquad \Theta_R(0) = \Theta_f(1), \tag{7}$$

$$\Theta_p(0) = \Theta_R(1); \qquad \Theta_f(0) = 0. \tag{8}$$

$\phi$ function is:

$$\phi(\alpha, \Theta_R) = Da(1-\alpha)^n \exp\left(\gamma \frac{\beta \Theta_R}{1 + \beta \Theta_R}\right). \tag{9}$$

The system (3)-(8) may be logistically written as

$$\Theta_{R(k+1)}(\xi_R) = F(\Theta_{Rk}(\xi_R)). \tag{10}$$

For the exemplary values of parameters $Da$=0.15, $n$=1, $\gamma$=10, $\beta$=1.4, $\delta$=2, $\delta_w$=0.5, $\sigma$=1, the bifurcation diagram is presented in Fig. 7. It may be seen that for $\Theta_H$=−0.09 the system has two stable steady states, the lower state and the upper one; thus neither of them can generate oscillations. However, due to the fact that the system is composed of two delaying elements (reactor and feedback), the system can generate flip-flop oscillations. In fact, introducing to the individual apparatuses adequate initial profiles, the time series of the reactor's outlet, $\Theta_R(1)$, has the shape as in Fig. 8. It may clearly be seen that lower and upper values of this time series correspond to the values of the lower and the upper steady state. Fig. 9 presents regions of initial conditions limited by of reactor's inlet $\Theta_R(0)$ and the inlet $\Theta_p(0)$. The shaded regions determine the generation of flip-flop oscillations. As results from Fig. 7, the value of $\Theta_H=-0.07$ destabilizes the upper branch of the bifurcation diagram, generating autonomous oscillations. It may clearly be seen that these are chaotic oscillations [4]. Adequate boundary conditions may bring about the generation of a semi-chaotic time series in the discussed reactor system, which is shown in Fig. 10. It is clearly seen that the lower values $\Theta_R(1)$ refer to the lower steady state. They are repeated at equal time intervals $\Delta\tau=2$. On the other hand, the upper values $\Theta_R(1)$ vary in a chaotic way. This may be evidenced by the sensitivity to the initial conditions, presented in the graph by the continuous line and the broken one.

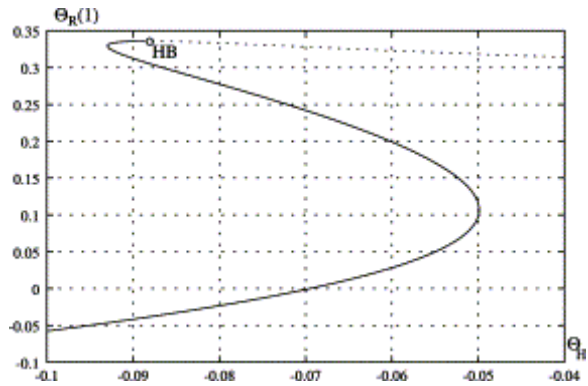

Fig. 7. Bifurcation diagram of a reactor.

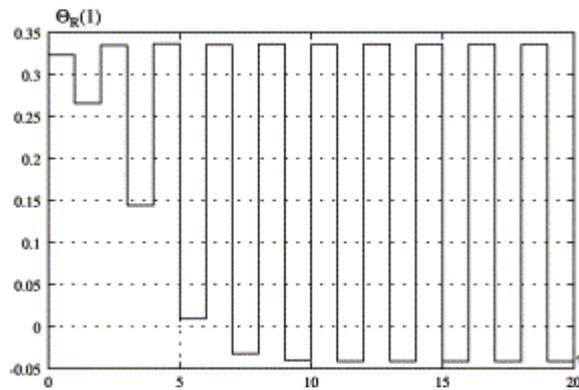

Fig. 8. Flip-flop oscillations in a reactor ($\Theta_H=-0.09$).

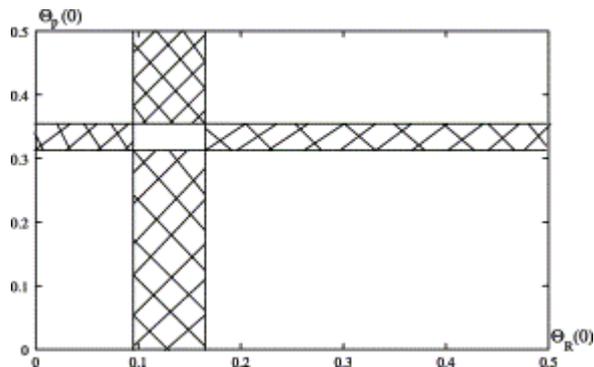

Fig. 9. Regions of initial conditions determining the flip-flop oscillations in a reactor ($\Theta_H=-0.09$).

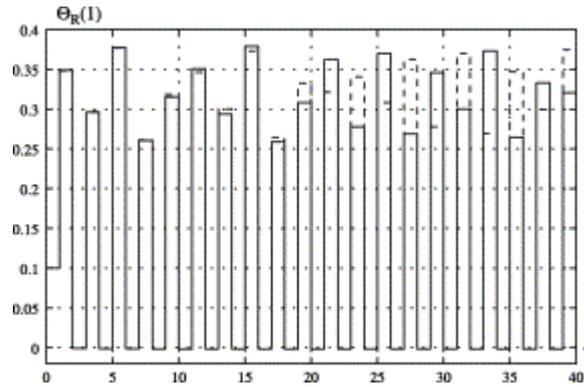

Fig. 10. Semi-chaotic flip-flop oscillations in a reactor ($\Theta_H = -0.07$).

# 5. Concluding remarks

In the paper specific type of oscillations, viz. flip-flop oscillations, are presented. They may be generated by a logistic system composed of at least two elements with distributed parameters. The mechanism of formation of these oscillations consists in a specific transmission of variables from one dynamic state to another. In chemical engineering this property is characteristic for all type of recirculation systems with delay. In particular this is true for reactors [3-6]. The above-mentioned oscillations may appear in the case of lack of dynamic bifurcations in the system. They are characterized by jumps of values of variables between individual steady states. The flip-flop oscillation may also appear in the presence of dynamic bifurcations. Their amplitudes are modulated in this case by an autonomous oscillation. In consequence, when autonomous oscillations are of chaotic type, the mentioned modulation is also a chaotic one. In consequence, the complete time series of a state variable may become a semi-chaotic series, i.e. fragments: regular and predictable as well as irregular and unpredictable.